\documentclass[aps,prb,twocolumn,floats]{revtex4}

\usepackage{ifthen}
\usepackage{ifpdf}
\usepackage{balance}

\usepackage{latexsym}
\usepackage{amsmath}
\usepackage{amssymb}
\usepackage{bm}
\usepackage{wasysym}

\ifpdf
\usepackage{graphicx}
\usepackage{epstopdf}
\else
\usepackage{graphicx}
\usepackage{epsfig}
\fi

\newcommand{\eexp}{\mbox{e}^}

\newcommand{\mass}{\mathsf{m}}

\newcommand{\tbox}[1]{\mbox{\tiny #1}}

\newcommand{\amatrix}[1]{\begin{matrix} #1 \end{matrix}}

\newcommand{\be}[1]{\begin{eqnarray}\ifthenelse{#1=-1}{\nonumber}{\ifthenelse{#1=0}{}{\label{e#1}}}}
\newcommand{\ee}{\end{eqnarray}}
\newcommand{\beq}{\begin{eqnarray}}
\newcommand{\eeq}{\end{eqnarray}}

\newcommand{\hide}[1]{}

\newcommand{\mpg}[2][1.0\hsize]{\begin{minipage}[b]{#1}{#2}\end{minipage}}

\newcommand{\nn}{\nonumber}

\begin{document} 

\title{Renormalization of the dephasing by zero point fluctuations}

\author{Swarnali Bandopadhyay$^{1,2}$ and Doron Cohen$^{1}$}

\affiliation{
\mbox{$^1$Department of Physics, Ben-Gurion University, Beer-Sheva 84105, Israel} \\
\mbox{$^2$Max Planck Institute for the Physics of Complex Systems, N{\"o}thnitzer Str. 38, 01187 Dresden, Germany}
}

\begin{abstract}
We study the role of zero-point-fluctuations (ZPF) in dephasing at
low temperature. Unlike the Caldeira-Leggett model where
the interaction is with an homogeneous fluctuating field of force, 
here we consider the effect of short range scattering by localized 
bath modes. We find that in presence of ZPF the inelastic
cross-section gets renormalized. Thus indirectly ZPF might contribute 
to the dephasing at low temperature.
\end{abstract}

\maketitle

\section{Introduction}\label{intro}

The coherent motion of a quantum mechanical particle in a fluctuating 
environment is endangered by decoherence due to inelastic scattering events.  
The temperature dependence of the resulting ``dephasing" effect 
has been worked out in numerous studies \cite{AAK,chak,imry}. 
During the last decade a controversy has emerged in the mesoscopic
literature regarding the role of zero-point-fluctuations (ZPF)
in the theory of low temperature dephasing.
The controversy was sparked by the experiment of Ref.\cite{webb} 
where a saturation of the dephasing rate in the limit 
of zero temperature has been reported, 
and consequently ZPF induced dephasing has been 
suggested as an explanation \cite{zaikin} and 
debated \cite{alt,golub,delft,dph}.


Possibly one can insist that ZPF lead to "${T=0}$" dephasing 
for a Brownian particle that interacts 
with an Ohmic Caldeira-Leggett (CL) bath \cite{CL} 
where the fluctuations of the environment consist 
of long wavelength (${q=0}$) modes. 
However \cite{dld,qbm}, in metallic environment the 
effective fluctuations are characterized 
by a finite correlation distance,
and hence consist of modes with wavenumbers~$q$
that range up to the Fermi momentum.
It was largely accepted \cite{dph,golub} that if the interactions
are short range, such that the fluctuations are
characterize by a finite correlation distance,
then the effect of the ZPF would be to renormalize
the scattering cross-section and the mass of the particle.
It turns out that in the cases of physical interest, 
and in particular for the prototype model of Refs.\cite{guinea,golubev},  
this renormalization effect is non-diverging in the zero
temperature limit: both mass renormalization \cite{horovitz} 
and dephasing \cite{rgd} for a single particle in the presence 
of a dirty metal environment have been studied.
Essentially the same formulation as in Refs.\cite{dph,rgd}
arise also in the more complicated many body treatment
of the dephasing problem \cite{munich}.


Though it is not diverging in problems of physical interest,
still the renormalization effect due to ZPF might be
significant in the actual analysis. The simplest 
possibility is to have an overall suppression 
of both elastic and inelastic scattering 
via a {\em Debye-Waller~factor} \cite{waller}. 
But we would like to explore the more exotic possibility 
of having a distinct enhancement ``factor" for the inelastic effect. 
It is therefore desired to have at least one exactly solvable
model for dephasing due to short range scattering
with environmental modes, that can be contrasted
with the opposite CL limit where all 
the mode have $q=0$. 
The objective of the present paper is to present such a model:
In the proposed model (see Fig.~1) the environment consists of
infinitely many localized fluctuating modes 
with (say) Ohmic spectral function, while the interaction 
is of short range and described by $\delta(x)$ as in ``s-scattering". 
This should be contrasted with the long range interaction 
of the CL model which is linear in $x$.


The outline of this paper is as follows: 
In Section~2 we define a model for a 
localized bath that induces 
both {\em zero point fluctuations} (ZPF) 
and {\em thermal fluctuations} (TRF).
In Sections~3-4 explicit expression for its 
scattering matrix are derived following \cite{thesis}.
In Section~5-6 low temperatures are considered,    
where the TRF are treated as a small perturbation. 
The example and the numerical analysis in Sections~7-8 
establish that for weak TRF 
the effect of the ZPF background 
can be taken into account by defining 
a renormalized intensity of the TRF.  
Accordingly ZPF may contribute to the dephasing at
low temperatures, though not directly.

\section{The Model}

The Hamiltonian of the particle plus the local bath is 
\be{0}
\mathcal{H} = \frac{p^2}{2\mass} 
+ \delta(x) \sum_{\alpha} c_{\alpha} Q_{\alpha} + 
\sum_{\alpha}\hat{n}_{\alpha}\omega_{\alpha}
\label{hamiltonian}
\ee
The index $n_{\alpha}=0,1,2,3...$ may  
indicate the state of the ${\alpha}$ oscillator, 
or optionally $n_{\alpha}=0,1$ may indicate the state 
of a two level (``spin") entity, as in our numerics.
From now on we use the notation
\be{0}
\hat{Q} = \sum_{\alpha} c_{\alpha}Q_{\alpha}
\label{interaction}
\ee
Assuming an incident particle with kinetic energy $\epsilon_k$ 
we divide the oscillators into two groups: 
those with $\omega_{\alpha}<\epsilon_k$ and those 
with $\omega_{\alpha}>\epsilon_k$ respectively.
We further assume low temperatures such that 
all the oscillators in the latter group 
are in the ground state. Note that the particle 
has enough energy to induce real (non-virtual) excitation 
of any of the TRF oscillators.    
Hence we can write schematically:
\beq
\hat{Q}  =  
\ \ c_{\tbox{S}} 
+ \sum_{\alpha\in\mbox{ZPF}} c_{\alpha} {Q_{\alpha}}
+ \sum_{\alpha\in\mbox{TRF}} c_{\alpha} {Q_{\alpha}} 
\label{Qscheme}
\eeq
where $c_{\tbox{S}}$ represents a static scatterer.
The particle is affected by the fluctuations of~$Q$. 
Assuming that the bath is prepared in the state $n=m$ 
the fluctuations are characterized by 
the non-symmetrized power spectrum 
\beq
\tilde{S}(\omega) 
= \overline{\sum_{n (\ne m)} 
|Q_{nm}|^2 \, 2\pi \delta\left(\omega-(E_n-E_{m})\right)}
\label{Sw}
\eeq
In the next section we show how the ZPF oscillators can be 
eliminated, such that the interaction is characterized 
by a dressed interaction matrix $\mathcal{Q}$.  
Accordingly we define the effective power spectrum as 
\beq
\tilde{S}_{\tbox{eff}}(\omega) 
= \overline{\sum_{n (\ne m)} 
|\mathcal{Q}_{nm}|^2 \, 2\pi \delta\left(\omega-(E_n-E_{m})\right)}
\label{Seff}
\eeq
and the {\em effective} ``size" of the elastic scatterer as 
\be{0}
c_{\tbox{eff}} = \mathcal{Q}_{m,m}
\ee  
In the following sections we explain how to define $\mathcal{Q}$ 
and how to make the exact calculation of 
the elastic scattering amplitude $\mathcal{T}$, 
and of the inelastic scattering cross section $p_{\tbox{inelastic}}$ (see Fig.~1). 
Then we discuss whether the results can be deduced from the effective
values of $c_{\tbox{eff}}$ and $\tilde{S}_{\tbox{eff}}(\omega)$.

\section{The Scattering states}

Outside of the scattering region the total energy 
of the system (particle plus bath) is 
\be{0}
\mathcal{E} \ \ = \ \ \epsilon_k + E_{n} 
\ \ = \ \ 
\epsilon_k + \sum_{\alpha}n_{\alpha}\omega_{\alpha}
\label{propcondn}
\ee
We look for scattering states that satisfy the equation
\[
\mathcal{H}|\Psi\rangle=\mathcal{E}|\Psi\rangle 
\]
Open (propagating) channels are those for which ${\epsilon_k >0}$ after scattering.
Otherwise the channels are closed (evanescent).
The channels are labeled as 
\be{0}
\bm{n} &=& (n_0,n) =  (n_0,n_{\tbox{ZPF}},n_{\tbox{TRF}}) 
\nn\\
&=& (n_{0} , n_{1},n_{2},n_{3},...,n_{\alpha},...)
\label{chanlindx}
\ee
where $n_0=\mbox{\footnotesize L,R}$ for left/right,  
and $n_{\tbox{ZPF}},n_{\tbox{TRF}}$ are collective 
indexes for the two group of scatterers. We define
\beq
k_{\bm{n}} =& \sqrt{2m(\mathcal{E}-E_n)}
\ \ \ \ \ & \mbox{for $n\in$ open}
\\ 
\alpha_{\bm{n}} =& \sqrt{-2m(\mathcal{E}-E_n)}
\ \ \ \ \ & \mbox{for $n\in$ closed}
\eeq
later we use the notations
\beq
v_{\bm{n}} &=&  k_{\bm{n}} / \mass \\
u_{\bm{n}} &=&  \alpha_{\bm{n}} / \mass
\eeq
and define diagonal matrices $\bm{v}=\mbox{diag}\{v_n\}$ 
and $\bm{u}=\mbox{diag}\{u_n\}$. 
The channel radial functions are written as 
\beq
R(r) =& A_n \eexp{-ik_nr} + B_n \eexp{+ik_nr}  
\ \ \ \ & \mbox{$n\in$ open} \\
R(r) =& C_n \eexp{-\alpha_n r} 
\ \ \ \ & \mbox{$n\in$ closed}
\eeq
where $r=|x|$. 
The wavefunction can be written as 
\beq
\Psi(r, n_0, Q) = \sum_{n} R_{n_0,n}(r) \chi^{n}(Q)
\eeq 
The matching equations are
\beq
\Psi(0,\mbox{right},Q) - \Psi(0,\mbox{left},Q) &=& 0 \\
\frac{1}{2\mathsf{m}}
\left[\Psi'(0,\mbox{right},Q) + \Psi'(0,\mbox{left},Q)\right] 
&=& \hat{Q} \Psi(0,Q) \ \ \ 
\eeq
The operator $\hat{Q}$ is represented 
by the matrix $Q_{nm}$ that has the block structure 
\beq
Q_{nm}
= \left(\amatrix{ 
Q_{vv} & Q_{vu} \\  
Q_{uv} & Q_{uu}
} \right)
\eeq
The matching conditions lead to the following set of matrix equations
\beq
A_{\tbox{R}} + B_{\tbox{R}} &=& A_{\tbox{L}} + B_{\tbox{L}}  \nn\\
C_{\tbox{R}} &=& C_{\tbox{L}} \nn\\
-i\bm{v} (A_{\tbox{R}}-B_{\tbox{R}} + A_{\tbox{L}}-B_{\tbox{L}}) &=& 2Q_{vv}(A_{\tbox{L}}+B_{\tbox{L}}) + 2Q_{vu}C_{\tbox{L}} \nn\\
-\bm{u}(C_{\tbox{R}} + C_{\tbox{L}}) &=& 2Q_{uv}(A_{\tbox{L}}+B_{\tbox{L}}) + 2Q_{uu} C_{\tbox{L}} \nn
\eeq
From here we get the matching equations 
that relate the ingoing and the outgoing amplitudes:
\beq
A_{\tbox{R}} + B_{\tbox{R}} &=& A_{\tbox{L}} + B_{\tbox{L}} \\
A_{\tbox{R}}-B_{\tbox{R}} + A_{\tbox{L}}-B_{\tbox{L}} &=& i2 (\bm{v})^{-1} \mathcal{Q} (A_{\tbox{L}}+B_{\tbox{L}}) 
\eeq
where the dressed interaction matrix is defined as 
\be{21}
\mathcal{Q} = Q_{vv} 
- Q_{vu}
\frac{1}{(\bm{u} + Q_{uu})} 
Q_{uv}
\ee
In the next section we deduce the $\bm{S}$ matrix from the above set of equations, 
and obtain explicit expressions for the elastic scattering amplitude 
and for the inelastic cross section.

\section{The $S$ matrix}

The unitary description of the scattering in terms 
of ingoing and out going probability currents 
requires to define the normalized ingoing and the 
outgoing amplitudes as ${\tilde{A_n} = \sqrt{v_n} A_n}$ 
and ${\tilde{B_n} = \sqrt{v_n} B_n}$. 
Consequently we defined a re-scaled version 
of the $Q_{nm}$ matrix as follows:
\be{0}
M_{nm}
=\left(\amatrix{ 
M_{vv} & M_{vu} \\  
M_{uv} & M_{uu}
} \right)
= \left(\amatrix{ 
\frac{1}{\sqrt{v}}Q_{vv}\frac{1}{\sqrt{v}} 
& \frac{1}{\sqrt{v}}Q_{vu}\frac{1}{\sqrt{u}} \\  
\frac{1}{\sqrt{u}}Q_{uv}\frac{1}{\sqrt{v}} 
&\frac{1}{\sqrt{u}}Q_{uu}\frac{1}{\sqrt{u}}
} \right)
\label{blkM}
\ee
We also define a corresponding reduced matrix
\beq
\mathcal{M}
= 
\frac{1}{\sqrt{\bm{v}}} 
\mathcal{Q} 
\frac{1}{\sqrt{\bm{v}}} 
=
M_{vv} 
- M_{vu}
\frac{1}{1 + M_{uu}} 
M_{uv}
\label{Meff}
\eeq
Using these notations the set of matching conditions
can be expressed using a transfer matrix as follows:
\beq
\left( \amatrix{ \tilde{B}_{\tbox{R}} \\ \tilde{A}_{\tbox{R}} } \right) 
= \bm{T} \left(\amatrix{ \tilde{A}_{\tbox{L}} \\ \tilde{B}_{\tbox{L}} } \right) 
\eeq
The transfer $2N \times 2N$ matrix can be  
written in block form as follows:
\beq
\bm{T}
=
\left( \amatrix{ 
\bm{T}_{++} & \bm{T}_{+-} \\
\bm{T}_{-+} & \bm{T}_{--}      
} \right)
=
\left( \amatrix{ 
1-i\mathcal{M} & -i\mathcal{M} \\
i\mathcal{M} & 1 + i\mathcal{M}      
} \right)
\eeq
The $\bm{S}$ matrix is defined via 
\beq
\left( \amatrix{ \tilde{B}_{\tbox{L}} \\ \tilde{B}_{\tbox{R}} } \right) 
= \bm{S} \left( \amatrix{ \tilde{A}_{\tbox{L}} \\ \tilde{A}_{\tbox{R}} } \right) 
\eeq
and can be written in block form as 
\beq
\bm{S}_{\bm{n},\bm{m}} \ \ = \ \  
\left(\amatrix{ \bm{S}_R & \bm{S}_T \\ \bm{S}_T & \bm{S}_R } \right)
\eeq
A straightforward elimination gives 
\beq
\bm{S} \ \ = \ \ 
\left( \begin{array}{ccc} 
-\bm{T}_{--}^{-1}\bm{T}_{-+} 
& \ & 
\bm{T}_{--}^{-1} 
\\ 
\bm{T}_{++} {-} \bm{T}_{-+}\bm{T}_{--}^{-1}\bm{T}_{+-} 
& \ & 
\bm{T}_{+-} \bm{T}_{--}^{-1} 
\end{array}\right) 
\eeq 
Now we can write expressions for $\bm{S}_R$ and for $\bm{S}_T$ 
using the $\mathcal{M}$ matrix. 
\beq
\bm{S}_T &=& \frac{1}{1+i\mathcal{M}} 
= 1 - i\mathcal{M} - \mathcal{M}^2 +i\mathcal{M}^3 + ... \\
\bm{S}_R&=& \bm{S}_T - \bm{1}
\eeq

The elastic forward scattering amplitude is 
\be{0}
\mathcal{T} \ \ = \ \ [\bm{S}_T]_{\bm{m},\bm{m}} 
\ \ = \ \ \left[\frac{1}{1+i\mathcal{M}}\right]_{\bm{m},\bm{m}} 
\label{elsT}
\ee
The total elastic scattering probability is 
\be{0}
p_{\mbox{elastic}} 
& = &  
|\mathcal{T}|^2 + |\mathcal{T}-1|^2\nn \\
& = &
1 - 2\left[\Re(\mathcal{T}) - |\mathcal{T}|^2 \right]
\label{peltot}
\ee
We observe that the inelastic scattering 
is isotropic and its line shape (per direction) is 
\be{33}
p(\omega) 
=
\sum_{\bm{n}(\ne \bm{m})}  
|[\bm{S}_T]_{\bm{n}\,\bm{m}}|^2 
\,2\pi\delta(\omega-(E_{\bm{n}}{-}E_{\bm{m}}))
\ee
with the measure $d\omega/(2\pi)$. 
The total inelastic cross section is obtained by integration
\be{0}
p_{\mbox{inelastic}} 
\, &=& \, 
2\int \frac{d\omega}{2\pi} p(\omega)\nn
\\
\, &=& \, 
2\sum_{\bm{n}(\ne \bm{m})}  |[\bm{S}_T]_{\bm{n},\bm{m}}|^2\nn\\
\, &=& \,  
2\sum_{\bm{n}(\ne \bm{m})} 
\left|\left[\frac{1}{1+i\mathcal{M}}\right]_{\bm{n},\bm{m}}\right|^2 
\label{pinel}
\ee
One can verify the $p_{\mbox{inelastic}}$ and $p_{\mbox{elastic}}$ sum up to unity, 
which is essentially the ``optical theorem".

\section{Perturbation theory}

Since the temperature is low we treat 
the small effect of the TRF in leading order.
We write 
\be{0}
\hat{Q} &=&
\hat{Q}^{\tbox{ZPF}} \otimes  \bm{1}^{\tbox{TRF}} 
+\bm{1}^{\tbox{ZPF}} \otimes  \hat{Q}^{\tbox{TRF}}
\nn\\
&\equiv& \ \ \hat{Q}^0 + \delta \hat{Q} 
\ee 
where $\hat{Q}^{0}$ is the sum over the ZPF coordinates  
including the static scatterer~$c_{\tbox{S}}$,  
while $\delta \hat{Q}$ is the sum over the TRF coordinates. 
For the reduced $\mathcal{Q}$ matrix we get:  
\be{0}
&&
\hspace*{-3mm}
\mathcal{Q} = 
Q_{vv}^0 + \delta Q_{vv}
\\\nn
&&-Q_{vu}^0
\left(
\frac{1}{\bm{u}+Q_{uu}^0}-\frac{1}{\bm{u}+Q_{uu}^0} \delta Q_{uu} \frac{1}{\bm{u}+Q_{uu}^0}
\right)
Q_{uv}^0
\ee
We assume that all the ``important" open modes are well 
above the the evanescent threshold. This means 
that a single TRF transition is not enough to 
push the scattered particle into an evanescent mode.  
Accordingly $\delta Q_{uv}$ and $\delta Q_{vu}$ 
are not included. With the same spirit 
we further assume that the TRF transitions hardly affect 
the evanescent velocity, hence 
\be{0}
u_{n_{\tbox{ZPF}},n_{\tbox{TRF}}} \approx u_{n_{\tbox{ZPF}},0} 
\ee
When calculating the matrix element $\mathcal{Q}_{\bm{n}\bm{m}}$ 
the second term constitutes a sum over sequences 
${ Q^0_{n,\nu}...(\delta Q)_{\nu',\mu'}...Q^0_{\mu,m}}$. 
In order to have a non zero term, 
the TRF-oscillators of the $\nu$ state should 
remain in the same state as in the $n$ state, 
while one ZPF oscillator of the $\nu$ state has to be excited. 
Similar observation applies to the states 
of the oscillators of the $\mu$ state. 
The TRF transitions are induced by $\delta Q$ 
during the evanescent motion of the particle.
Accordingly we deduce that 
\be{0}
&&
\hspace*{-3mm}
\mathcal{Q} = 
\left(\left[Q^{\tbox{ZPF}}_{vv}-Q^{\tbox{ZPF}}_{vu}\left(\frac{1}{\bm{u}+Q^{\tbox{ZPF}}_{uu}}\right)Q^{\tbox{ZPF}}_{uv}\right]_{m,m}\right)
\bm{1}^{\tbox{TRF}} 
\nn\\
&&+\left(1+\left[Q^{\tbox{ZPF}}_{vu}\left(\frac{1}{\bm{u}+Q^{\tbox{ZPF}}_{uu}}\right)^2Q^{\tbox{ZPF}}_{uv}\right]_{m,m}\right)
Q^{\tbox{TRF}}
\nn
\ee
which can be written schematically as follows   
\be{38}
\mathcal{Q} \ \ = \ \ c_0\bm{1}^{\tbox{TRF}} \ + \ \lambda_0 Q^{\tbox{TRF}}
\ee
where the effective elastic scattering amplitude, 
and the scaling factor of the inelastic effect are    
\be{39}
c_0 &=& 
c_{\tbox{S}}- \left[Q^{\tbox{ZPF}}_{vu}\left(\frac{1}{\bm{u}+Q^{\tbox{ZPF}}_{uu}}\right)Q^{\tbox{ZPF}}_{uv}\right]_{m,m}
\\
\label{e40}
\lambda_0 &=& 
1+\left[Q^{\tbox{ZPF}}_{vu}\left(\frac{1}{\bm{u}+Q^{\tbox{ZPF}}_{uu}}\right)^2Q^{\tbox{ZPF}}_{uv}\right]_{m,m}
\ee
With an appropriate counter term we can make ${c_0=0}$. 
More interestingly we see that the effective TRF are characterized 
by the dressed power spectrum 
\be{0}
\tilde{S}_{\tbox{eff}}(\omega) \ \ = \ \  (\lambda_0)^2 \, \tilde{S}(\omega)
\ee

\section{The dressed Born approximation}

The first order (``Born") approximation relates 
the inelastic line shape to the 
power spectrum of the fluctuations. 
We use the term ``dressed Born approximation" 
in order to indicate that we use 
first order perturbation theory with respect 
to the TRF, while the ZPF including 
the static scatterer are treated to infinite order.
Within this framework the leading order expression 
for the $\bm{S}$ matrix, using Eqs.(\ref{e38}-\ref{e40}), is 
\be{0}
\bm{S}_{T}
&\approx& 
\frac{1}{1+i(c_0/v_{\epsilon})+i\lambda_0\mathcal{M}^{\tbox{TRF}}} \nn\\
&=&  
\mathcal{T}_0\bm{1} - i \mathcal{T}_0^2 \lambda_0  \mathcal{M}^{\tbox{TRF}} + ...
\label{prtrbST}
\ee
where the elastic forward scattering amplitude is 
\be{0}
\mathcal{T}_0 = \frac{1}{1+i(c_0/v_{\epsilon})}
\label{teff}
\ee
Consequently we get for the inelastic scattering
\be{44}
p(\omega) 
\ \ \approx \ \ 
\frac{1}{v_{\epsilon}v_{\epsilon{-}\omega}} 
\, |\mathcal{T}_0|^4 \, (\lambda_0)^2 \,\tilde{S}(\omega)
\label{scalng}
\ee 
Since we had assumed that the change in the kinetic energy 
of the particle due to TRF inelastic scattering is relatively small, 
one can take $v_{\epsilon{-}\omega} \approx v_{\epsilon}$.

\section{The simplest example}

Consider a particle with velocity $v_{\epsilon}$, 
that collides with a `bath' that consists of an elastic 
scatterer $c_{\tbox{S}}$, and a single two level TRF scatterer $c_{\tbox{T}}$ 
whose excitation energy is $\omega_{\tbox{T}} (\ll \epsilon_k)$. 
The interaction matrix is  
\be{50}
\mathcal{Q} = Q = 
\left(\amatrix{
c_{\tbox{S}} & c_{\tbox{T}} \\
c_{\tbox{T}} & c_{\tbox{S}} } \right) 
\ee
which we substitute in $\mathcal{M} \approx (1/v_{\epsilon})^2 \mathcal{Q}$, 
so as to get ${\bm{S}_T=(1+i\mathcal{M})^{-1}}$. In order to avoid 
crowded expressions we set the units such that ${v_{\epsilon}=1}$, 
and write
\be{0}
\bm{S}_T =
\frac{1}{(1+ic_{\tbox{S}})^2+c_{\tbox{T}}^2}
\left(\amatrix{
1+ic_{\tbox{S}} & -ic_{\tbox{T}} \\
-ic_{\tbox{T}} & 1+ic_{\tbox{S}} } \right) 
\ee
[Note again that in order to restore the units 
each~$c$ in the above expression should be replaced 
by $c/v_{\epsilon}$]. From here it follows that 
\be{52}
p_{\tbox{inelastic}} = \frac{2\nu_{\tbox{TRF}}}{(1-c_0^2+\nu_{\tbox{TRF}})^2+4c_0^2}
\ \ \ \ \ \ \ \ \mbox{\small [scaled]}
\ee
where $\nu_{\tbox{TRF}}\equiv c_{\tbox{T}}^2$ characterizes 
the intensity of the TRF, and $c_0 \equiv c_{\tbox{S}}$. 
One observes that for strong TRF the inelastic effect 
is suppressed and we get mainly elastic back reflection. 
But in the regime of interest, of weak TRF,  
the inelastic scattering is proportional to $\nu{\tbox{TRF}}$ 
and agree with Eq.(\ref{e44}) 
where ${|\mathcal{T}_0|^2 = 1/(1{+}c_0^2)}$ and ${\lambda_0=1}$.

\hide{
\be{0}
|\mathcal{T}_0|^2 \ \ = \ \ \frac{1}{1+c_0^2}
\ \ \ \ \ \ \ \ \mbox{\small [scaled]}
\ee
}

Next we complicate the `bath' by adding 
a single ZPF scatterer $c_{\tbox{Z}}$ 
whose excitation energy is $\omega_{\tbox{Z}} (> \epsilon_k)$.
The possible values of the mode index 
are $n=(0,0) \equiv 1$, 
and $n=(0,1) \equiv 2$, 
and $n=(1,0) \equiv 3$, 
and $n=(1,1) \equiv 4$.
The ZPF scatterer is assumed to be 
in the ground state ($m=1$),    
and hence only the first two modes are open. 
The interaction matrix is  
\be{0}
Q = \left(\amatrix{ 
c_{\tbox{S}} & c_{\tbox{T}} & c_{\tbox{Z}} & 0 \\
c_{\tbox{T}} & c_{\tbox{S}} & 0 & c_{\tbox{Z}} \\
c_{\tbox{Z}} & 0 & c_{\tbox{S}} & c_{\tbox{T}} \\
0 & c_{\tbox{Z}} & c_{\tbox{T}} & c_{\tbox{S}} 
} \right)\,.\nn
\ee
If we did not have the TRF oscillator, 
it would be a $2\times2$ matrix:
\be{0}
Q^{\tbox{ZPF}} =   
\left(\amatrix{
c_{\tbox{S}} &  c_{\tbox{Z}}\\
c_{\tbox{Z}} & c_{\tbox{S}}} \right)
\ee
If we ignored the ZPF oscillator, 
we would get Eq.(\ref{e50}). 
But using Eq.(\ref{e21}) we get 
the dressed interaction matrix: 
\be{0}
\mathcal{Q}&=&\left(\amatrix{
c_{\tbox{S}} & c_{\tbox{T}} \\
c_{\tbox{T}} & c_{\tbox{S}} } \right)\, 
\nn\\
&-& \frac{c_{\tbox{Z}}^2}{(u_3{+}c_{\tbox{S}})\,(u_4{+}c_{\tbox{S}})-c_{\tbox{T}}^2}\,
\left(\amatrix{
(u_3{+}c_{\tbox{S}}) & -c_{\tbox{T}} \\
-c_{\tbox{T}} & (u_4{+}c_{\tbox{S}}) } \right)
\nn
\ee
with $u_3=\sqrt{|\epsilon_k-\omega_{\tbox{Z}}|}$ 
and $u_4=\sqrt{|\epsilon_k-\omega_{\tbox{Z}}-\omega_{\tbox{T}}|}$.
Consequently from ${c_{\tbox{eff}} \equiv \mathcal{Q}_{1,1}}$ we get:
\be{0}
c_{\tbox{eff}} \ \ = \ \  c_{\tbox{S}}-
\frac{(u_3{+}c_{\tbox{S}}) \, \nu_{\tbox{ZPF}}}{(u_3{+}c_{\tbox{S}})\,(u_4{+}c_{\tbox{S}})-\nu_{\tbox{TRF}}}
\label{cefFulEx2}
\ \ \ 
\ee
and from $\nu_{\tbox{eff}} \equiv |\mathcal{Q}_{2,1}|^2$  
we get $\nu_{\tbox{eff}} = \lambda^2 \nu_{\tbox{TRF}}$ where 
\be{51}
\lambda \ \ = \ \ 
1+\frac{\nu_{\tbox{ZPF}}}{(u_3{+}c_{\tbox{S}})\,(u_4{+}c_{\tbox{S}})-\nu_{\tbox{TRF}}} 
\label{nuefFulEx2}
\ \ \ 
\ee
with $\nu_{\tbox{ZPF}}\equiv c_{\tbox{Z}}^2$ 
and $\nu_{\tbox{TRF}}\equiv c_{\tbox{T}}^2$. 
Optionally we can get for $\mathcal{Q}$ the approximated result 
of Eq.(\ref{e38}), which treats the TRF coupling in leading order. 
This treatment assumes that 
in the vicinity of the energy shell   ${v_1 \approx v_2 \equiv  v_{\epsilon}}$, 
and ${u_3 \approx u_4 \equiv  u_{\tbox{Z}} }$.  
The parameters $c_0$ and $\lambda_0$ are calculated 
using Eqs.(\ref{e39}-\ref{e40}) with $Q^{\tbox{ZPF}}_{uu}=c_{\tbox{S}}$, 
and $Q^{\tbox{ZPF}}_{vu}= Q^{\tbox{ZPF}}_{uv} = c_{\tbox{Z}}$, 
leading to 
\be{0}
c_0 &=& 
c_{\tbox{S}} -\frac{\nu_{\tbox{ZPF}}}{(u_{\tbox{Z}}+c_{\tbox{S}})}
\label{ceffEx2}
\\
\lambda_0 &=& 
1+\frac{\nu_{\tbox{ZPF}}}{(u_{\tbox{Z}}+c_{\tbox{S}})^2}
\label{lmbdEx2}
\ee
In this simple example the dependence of $c_0$
and $\lambda_0$ on $\nu_{\tbox{ZPF}}$ is linear.
But once we have more than one ZPF scatterer (as in the 
numerical example of the next section) the relation is 
no longer linear. It might be also of interest to solve 
the first equation ${c_0=0}$ for $c_{\tbox{S}}$,  
and substitute the result into the second equation. 
The outcome of this procedure is illustrated in Fig.~3.

The calculation of the $\bm{S}$ matrix proceed in the same 
way as in the single TRF case, with the effective 
interaction matrix (no approximation involved):
\be{0}
\mathcal{Q} = 
\left(\amatrix{
c_{\tbox{S}}-(\lambda{-}1)(u_3{+}c_{\tbox{S}}) & \lambda c_{\tbox{T}} \\
\lambda c_{\tbox{T}} &  c_{\tbox{S}}-(\lambda{-}1)(u_4{+}c_{\tbox{S}})} \right) 
\ee
Setting ${v_1 \approx v_2 \equiv  v_{\epsilon}}$ 
and ${u_3 \approx u_4 \equiv  u_{\tbox{Z}} }$  as before, 
we label both diagonal terms as $c_{\tbox{eff}}$. 
Still we are not making any approximation with regard 
to the intensities $\nu_{\tbox{TRF}}$ and $\nu_{\tbox{ZPF}}$, 
so as to get essentially exact results:
\be{55}
|\mathcal{T}|^2 \ \ = \ \ 
\frac{[v_{\epsilon}^2+(c_{\tbox{eff}})^2] \, v_{\epsilon}^2} 
{[v_{\epsilon}^2  - (c_{\tbox{eff}})^2  + \lambda^2 \nu_{\tbox{TRF}}]^2 + 4v_{\epsilon}^2(c_{\tbox{eff}})^2}
\ee
and the generalization of Eq.(\ref{e52}): 
\be{56}
p_{\tbox{inelastic}} \ \ = \ \ 
\frac{2}{v_{\epsilon}^2} \ |\tilde{\mathcal{T}}|^2 |\mathcal{T}|^2  \ \lambda^2  \ \nu_{\tbox{TRF}}
\ee
where $|\tilde{\mathcal{T}}|^2$ is Eq.(\ref{e55}) 
without the $\lambda^2 \nu_{\tbox{TRF}}$ term.
For weak TRF intensity, using $\lambda\approx\lambda_0$ 
and $\mathcal{T} \approx \mathcal{T}_0$  
one obtains the dressed Born approximation Eq.(\ref{e44}). 
One observes that the presence of the factor $\lambda$ 
has two implications: 
one is to enhance the inelastic scattering 
for weak TRF, while the other is to limit the range 
over which the weak TRF approximation applies.

\section{Discussion, Expectations, and Numerical demonstration}

The analysis in the present paper is focused primarily 
on the low temperature scattering, due to weak TRF, 
where the dressed Born approximation of Section~6 applies. 
Still in order to get the ``big picture" we consider  
below the full range of $\nu_{\tbox{TRF}}$ values. 
We first highlight some qualitative observations  
that are based on the analysis of the simple examples 
of the previous section, and then proceed with a 
numerical demonstration that involves a larger bath 
of scatterers.

From the Born approximation we deduce that for 
weak TRF the inelastic cross section $p_{\tbox{inelastic}}$  
is proportional to $\nu_{\tbox{TRF}}$. 
For strong TRF it drops down as implied e.g. 
by the simplest example Eq.(\ref{e52}). 
The maximum ${p_{\tbox{inelastic}}=1/2}$ 
is attained for the intermediate 
value ${\nu_{\tbox{TRF}}=c_0^2{+}1}$.
A-priory we could not expect a larger 
inelastic effect because the  
elastic cross-section ${|\mathcal{T}|^2+|\mathcal{T}{-}1|^2}$
is bounded from below by the minimum value $50\%$. 
We can interpret the condition for attaining minimum 
elastic cross section using a {\em Fabry-Perrot} 
double barrier picture: The elastic scattering  
and the inelastic scattering are like two barriers 
separated by an infinitesimal distance. 
The strongest interference effect is expected 
when the two barriers are comparable.

The suppression of the inelastic 
effect for strong TRF is a generic effect: 
it becomes almost obvious if we consider 
the scattering of a particle from 
a fluctuating region in a three dimensional space.  
In the latter context strong fluctuations would 
repel the particle from the scattering region, 
hence making inelastic excitations  
within the excluded volume less likely. 
So the strongest inelastic 
effect is experienced for intermediate 
values of $\nu_{\tbox{TRF}}$.

The inclusion of ZPF into the model renormalizes~$\nu_{\tbox{TRF}}$. 
The enhancement factor $\lambda$ is larger than unity (but finite) 
in the Born approximation limit, but if we go to very high temperatures 
(large $\nu_{\tbox{TRF}}$) this renormalization effect fades away and we 
get $\lambda=1$. See e.g. Eq.(\ref{e51}).
The crossover involves a wild variation of $\lambda$ (see Fig.~4), 
which implies that ${0<p_{\tbox{inelastic}}<1/2}$ 
goes through the whole range of possible values (Fig.~5).

It is important to point out that if the fluctuations had 
continuous (rather than discrete) power spectrum, 
the above described intermediate wild variation 
would be smoothed away. 
Thus in realistic circumstances we expect that also 
in the presence of ZPF the qualitative dependence 
of $p_{\tbox{inelastic}}$ on $\nu_{\tbox{TRF}}$ 
would be smooth, though renormalized 
by $\lambda_0$ at the limit of low temperatures.

For the numerical study we consider a bath that consists 
of two level scatterers. The energy splitting of the $\alpha$ scatterer  
is $\omega_{\alpha}$ and the interaction is described 
by the operator $Q_{\alpha}=${\tiny $\left(\amatrix{0 & 1 \cr 1 & 0}\right)$}. 
The strength of the interaction with the bath is characterized by the  
intensity of the fluctuations as obtained by 
integrating over their power-spectrum $\tilde{S}(\omega)$.
Consequently we distinguish between the intensity of 
the ZPF and the intensity of the TRF: 
\be{0}
\nu_{\tbox{ZPF}} &=& \sum_{\alpha \in ZPF}{c_\alpha^{2}} 
\label{nuC} 
\\
\nu_{\tbox{TRF}} &=& \sum_{\alpha \in TRF}{c_\alpha^{2}} 
\label{nuT}
\ee
The {\em effective} intensity of the thermal fluctuations 
is similarly defined and accordingly calculated from 
the dressed interaction matrix:
\be{0}
\nu_{\tbox{eff}} &=& \sum_{n (\ne m)} \,\Big|\mathcal{Q}_{n,m}\Big|^2
\ee

Given a set of $N$ ZPF-scatterers with couplings $c_{\tbox{Z}}$, 
and a static scatterer $c_{\tbox{S}}$,  we calculate $c_0$ 
(which determines~$\mathcal{T}_0$) and $\lambda_0$ 
as a function of ${\nu_{\tbox{ZPF}}\equiv N|c_{\tbox{Z}}|^2}$. 
See Fig.~6.
Then, for various values of $\nu_{\tbox{ZPF}}$  we calculate 
the exact results for $\nu_{\tbox{eff}}$ and $p_{\tbox{inelastic}}$ 
versus $\nu_{\tbox{TRF}}$. 
See Fig.~7.
One expects that 
for weak TRF the effective intensity $\nu_{\tbox{eff}}$ 
would be proportional to $\nu_{\tbox{TRF}}$, namely  
\be{0}
\nu_{\tbox{eff}} \ \ \approx \ \  (\lambda_0)^2 \ \nu_{\tbox{TRF}}
\label{nurelatn}
\ee
Furthermore our perturbative scheme implies that  
\be{49}
p_{\tbox{inelastic}}
\ \ \approx \ \ 
\frac{2}{v_{\epsilon}^2} \ |\mathcal{T}|^4 \ \nu_{\tbox{eff}}
\ee
In order to test the quality of the latter approximation  
we re-plot the results for $p_{\tbox{inelastic}}$ 
versus $|\mathcal{T}|^4 \nu_{\tbox{eff}}$. See Fig.~8.
The numerical results confirm our qualitative expectations, 
and are in agreements with the analysis of the simple example 
of the previous section. In particular one observes 
that the presence of ZPF has two implications: 
one is to enhance the inelastic scattering 
for weak TRF, while the other is to limit the range 
over which the weak TRF approximation applies.

\section{Summary}

In the Caldeira-Leggett model the effect of the environment 
is characterized by a friction coefficient~$\eta$ and by a temperature~$T$. 
But more generally \cite{dld,qbm,dph,rgd} it has been emphasized 
that the proper way to characterize the environment 
is by its form factor $\tilde{S}(q,\omega)$. 
The form factor contains information on both the temporal 
and the spatial aspects of the fluctuations, and
in particular one can extract from it not only $T$ and $\eta$, but
also the spatial correlations. The general formula for the rate 
of dephasing \cite{dph,rgd} involves a $dq d\omega$ integral 
over $\tilde{S}(q,\omega)$, and for short range interactions 
simply reflect the rate of inelastic events.

We find for our model system that the inelastic scattering 
cross-section $p_{\tbox{inelastic}}$ is enhanced in the presence of ZPF, 
and accordingly ZPF may contribute to the dephasing at low temperatures, 
though indirectly. This might come as a surprise since 
in Ref.\cite{waller} it has been argued that both elastic 
and inelastic scattering are suppressed by ZPF by 
the {\em same} Debye-Waller factor (DWF). 
A closer look reveals the difference between the two models involved. 
In Ref.\cite{waller} one considers the scattering 
of a particle ($\hat{x}$) from a vibrating scatterer ($\hat{Q}$), 
where the interaction is ${\delta(\hat{x}{-}\hat{Q})}$.   
Accordingly the particle experiences a fluctuating 
field ${\hat{\mathcal{U}}(x)=\delta(\hat{Q}-x)}$ 
and $\tilde{S}(q,\omega)$ is the Fourier transform  
of ${\langle \eexp{-iq\hat{Q}(t)}\eexp{iq\hat{Q}(0)}\rangle}$, 
which is suppressed by the DWF ${\eexp{-\langle \hat{Q}^2 q^2\rangle}}$.
In our model the interaction is ${\delta(\hat{x})\hat{Q}}$. 
Accordingly $\tilde{S}(q,\omega)$ is the Fourier transform 
of ${\langle\hat{Q}(t)\hat{Q}(0)\rangle}$, 
and, within the framework of the conventional Born approximation, 
there is no DWF involved: in our model adding high frequency components 
to the fluctuating field has no implication on the low frequency  
behavior of $\tilde{S}(q,\omega)$.

The renormalization factor of the inelastic effect ($\lambda$) 
in our {\em dressed} Born approximation comes from higher orders 
of perturbation theory with respect to the ZPF, while the TRF 
are treated in leading order. 
The renormalization factor $\lambda$ multiplies  
the power spectrum $\tilde{S}(\omega)$ that  
describes the thermal fluctuations. The power spectrum 
itself does not involve a DWF.
The $\lambda$ renormalization of the inelastic scattering 
comes ``on top" of the expected renormalization 
of the {\em potential~floor} and of the {\em inertial~mass}, 
which are familiar from the solution of the Polaron problem. 
In our scattering theory framework the expected renormalization 
of the potential floor can be deduced from  
the ZPF induced offset in the effective `size' 
of the elastic scatterer~($c_{\tbox{S}}$), 
while the renormalization of the mass comes from the associated 
energy dependence of the forward scattering amplitude~($\mathcal{T}$).

One can construct an extended bath that consists  
of an homogeneously distributed set of ``s-scatterers", 
as described in \cite{dld}. This would allow the modeling 
of a fluctuating environment of physical interest 
(say a Dirty metal environment) with the desired $\tilde{S}(q,\omega)$.   
In such physical circumstances we expect renormalization of 
{\bf (i)} the potential floor; {\bf (ii)} the inertial mass;  
and {\bf (iii)} the effective thermal fluctuations.
Our results imply that these renormalization effects 
are non-divergent if the fluctuations are characterized 
by short range spatial correlations, but still they might 
modify the low temperature dependence of the dephasing effect.


\noindent
{\bf Acknowledgment:} 
Part of the derivation in Sections 3-4 has been done 
in collaboration with Chen Sarig \cite{thesis}. 
DC thanks Ora Entin-Wohlman (BGU) and Joe Imry (Weizmann Inst.)  
for a discussion that had illuminated the significance 
of the model in the Debye-Waller perspective.
SB thanks Grigory Tkachov (MPIPKS) for his interest in this work.
The research has been supported by the the Israel Science Foundation (grant No.11/02),  
and by a grant from the DIP, the Deutsch-Israelische Projektkooperation.



\clearpage


\mpg{
\begin{center}
\includegraphics[clip,width=8cm]{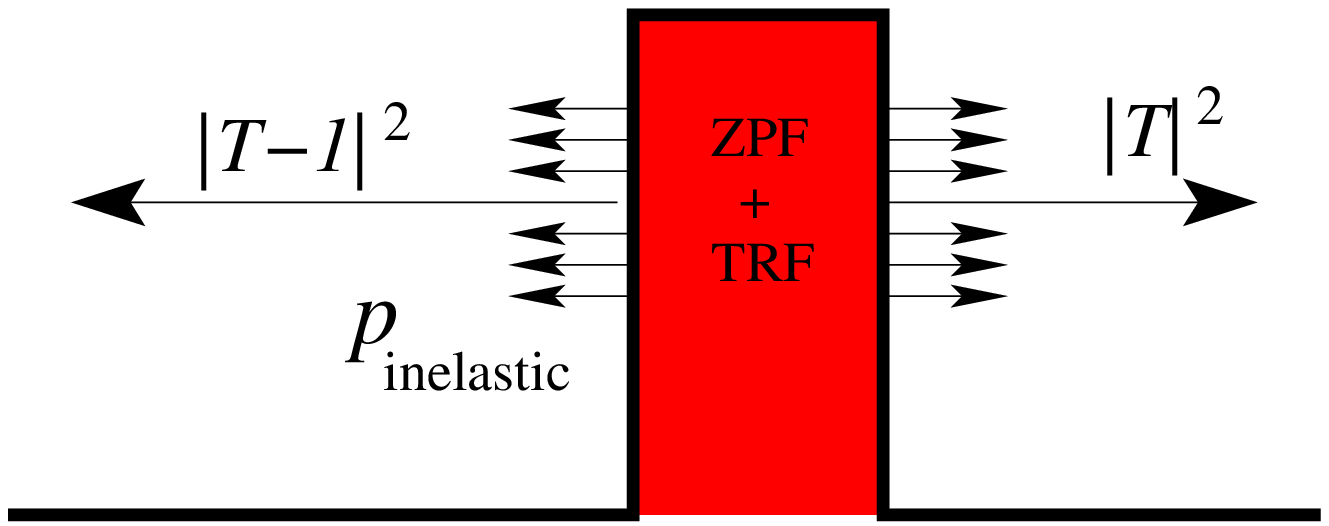}
\end{center}

{\footnotesize {\bf Fig.1:} 
Schematic diagram of the model system. The scattered 
wave of a particle that collides from the right 
with a thermal "s-scatterer" consists of forward 
elastic scattering with amplitude~$\mathcal{T}$, 
backward elastic scattering with amplitude~$\mathcal{T}{-}1$,
and isotropic inelastic scattering with probability $p_{\tbox{inelastic}}$. 
Our purpose is to find the dependence of  $\mathcal{T}$ 
and $p_{\tbox{inelastic}}$ on the intensity of the 
low temperature thermal fluctuations (TRF), with 
arbitrarily large background of zero point fluctuations (ZPF).  
}

}

\ \\

\mpg{
\begin{center}
\includegraphics[clip,width=8cm]{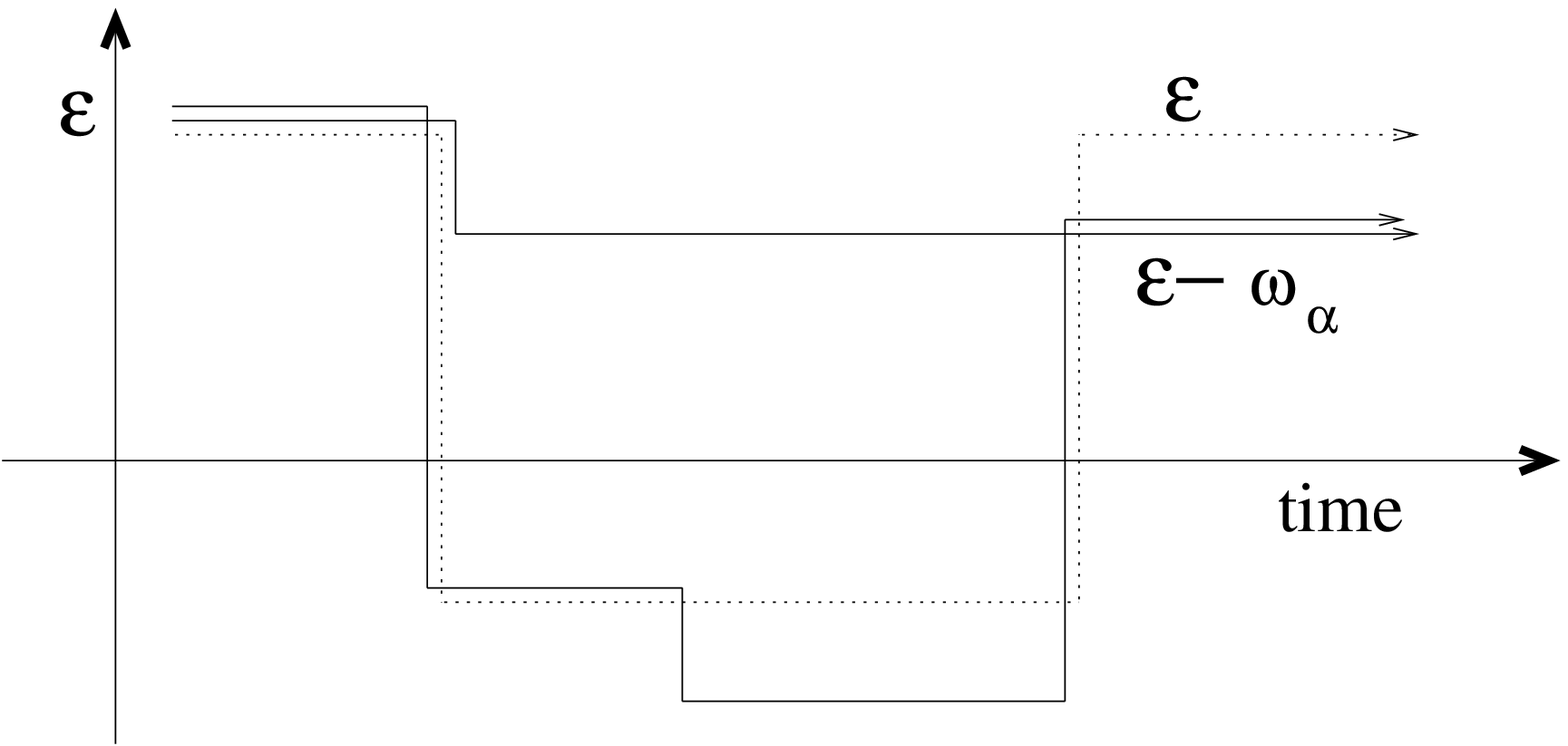}
\end{center}

{\footnotesize {\bf Fig.2:}
Diagrams that describe the time evolution 
of kinetic energy can be used in order to illustrate 
terms the scattering calculation.
The dotted line represents a contribution to the 
elastic cross section due to (virtual) scattering 
by ZPF modes. The solid lines represent contributions 
to the first order inelastic cross-section, where the 
intensity of the TRF is regarded as the small parameter.  
}

}

\ \\


\mpg{
\begin{center}
\includegraphics[clip,width=0.8\hsize]{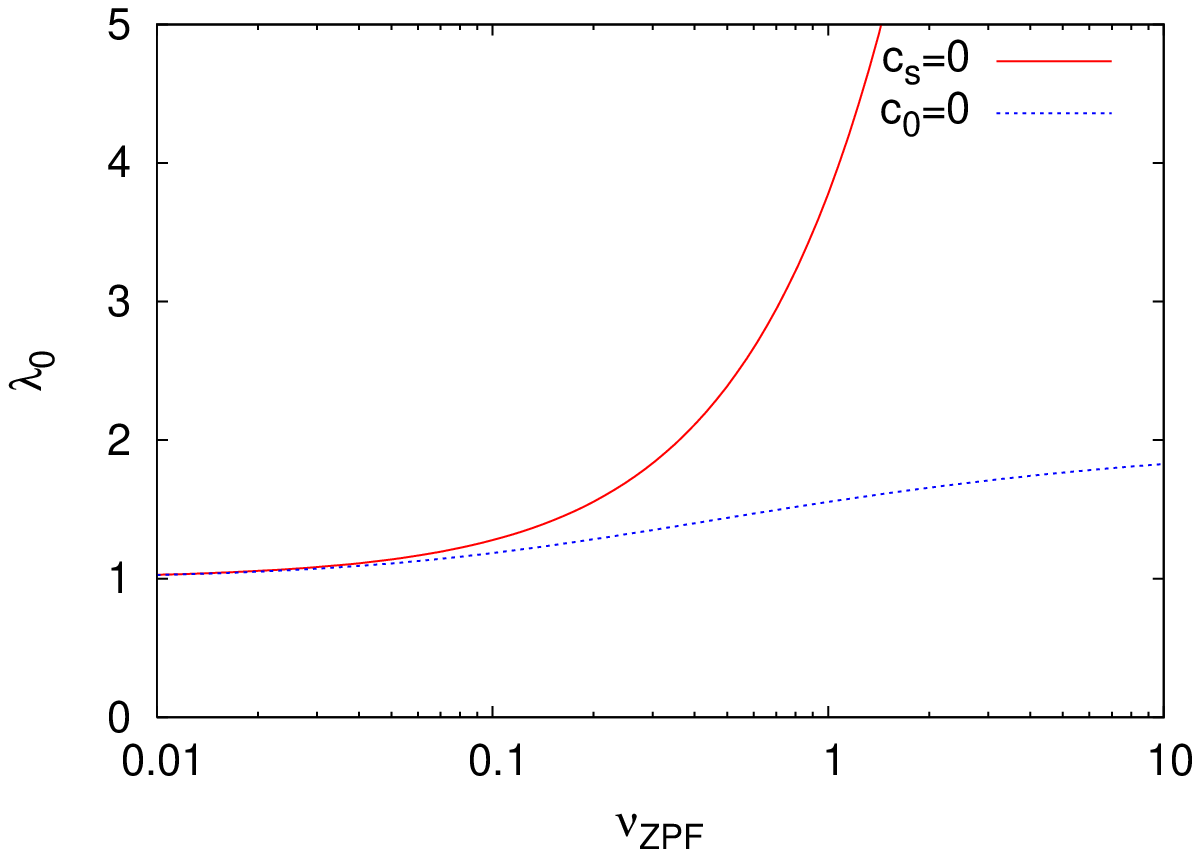}
\end{center}

{\footnotesize {\bf Fig.3:} 
The renormalization factor for inelastic effect $\lambda_{\tbox{ZPF}}$ 
is plotted as a function of $\nu_{\tbox{ZPF}}$ 
for the simple model of Section~7, using Eq.(\ref{lmbdEx2}) 
with fixed static scatterer ${c_{\tbox{S}}=0}$ (solid red curve) 
and with adaptive static scatterer such that $c_0=0$ (dashed blue curve).
The other parameters for this and for the next figures 
are ${\omega_{\tbox{Z}}=0.96}$ and ${\omega_{\tbox{T}}=0.03}$ and ${\epsilon_k=0.6}$.} 
}

\ \\

\mpg{
\begin{center}
\includegraphics[clip,width=0.75\hsize]{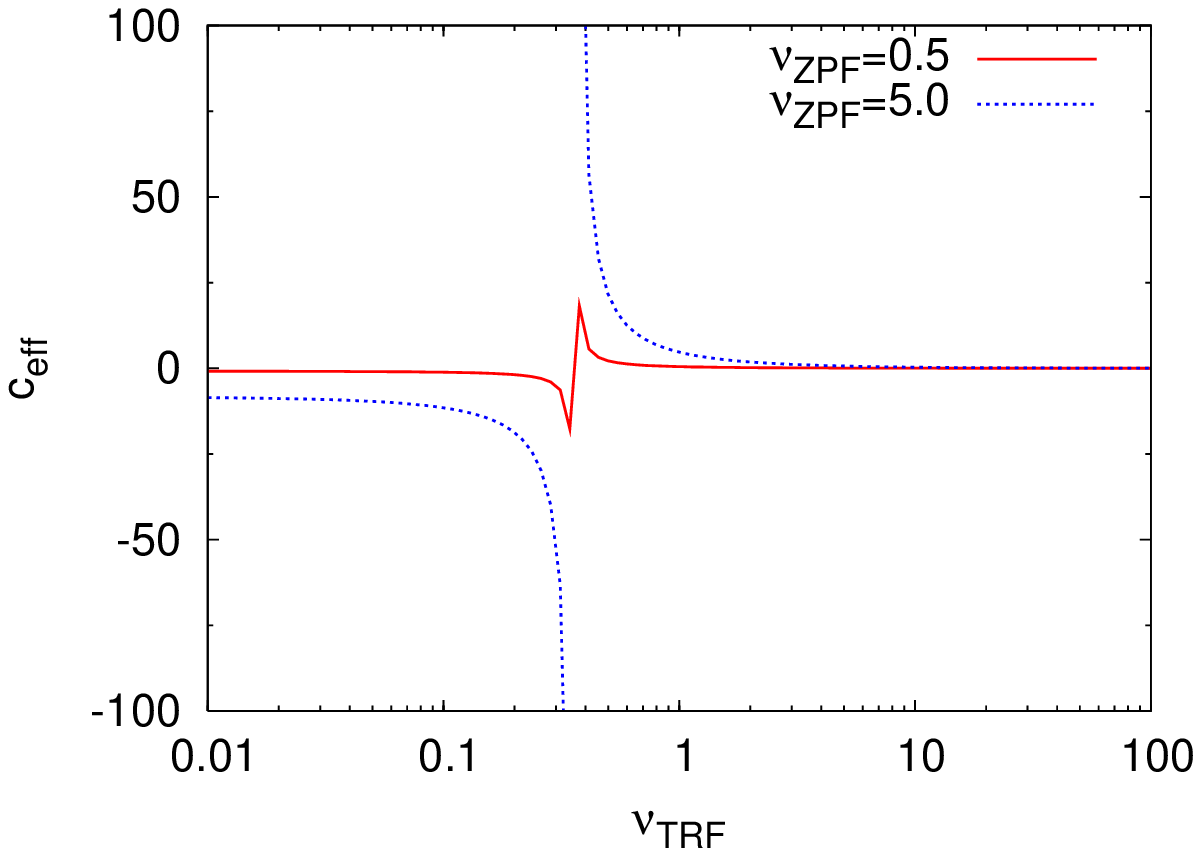} \\
\includegraphics[clip,width=0.75\hsize]{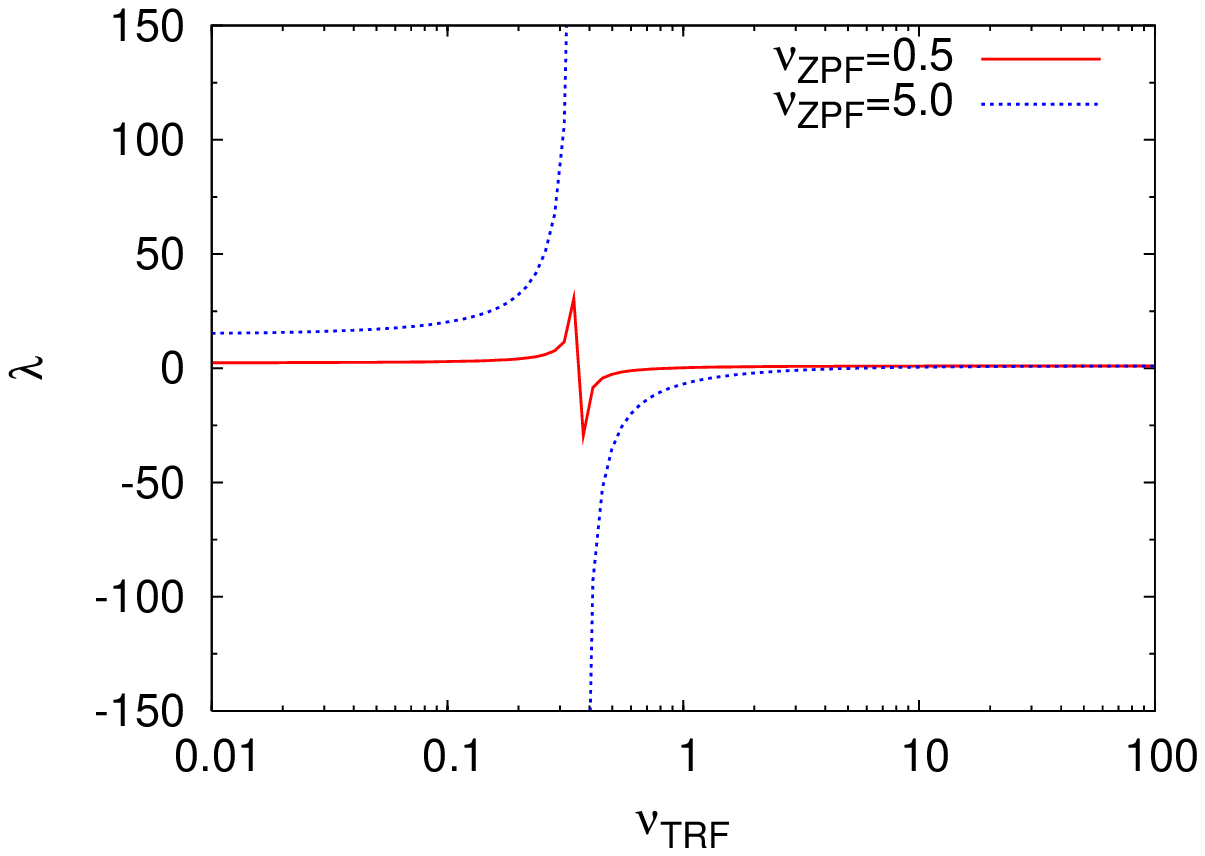}
\end{center}

{\footnotesize {\bf Fig.4:} 
Plots of ${c_{\tbox{eff}}}$ (upper panel) and $\lambda$ (lower panel) 
versus ${\nu_{\tbox{TRF}}}$ for the simple model of Section~7 
with the same parameters as in Fig.~3. The red solid curves are 
for ${\nu_{\tbox{ZPF}}=0.5}$ and the blue dashed curves are for ${\nu_{\tbox{ZPF}}=5}$.} 
}

\ \\

\mpg{
\begin{center}
\includegraphics[clip,width=0.75\hsize]{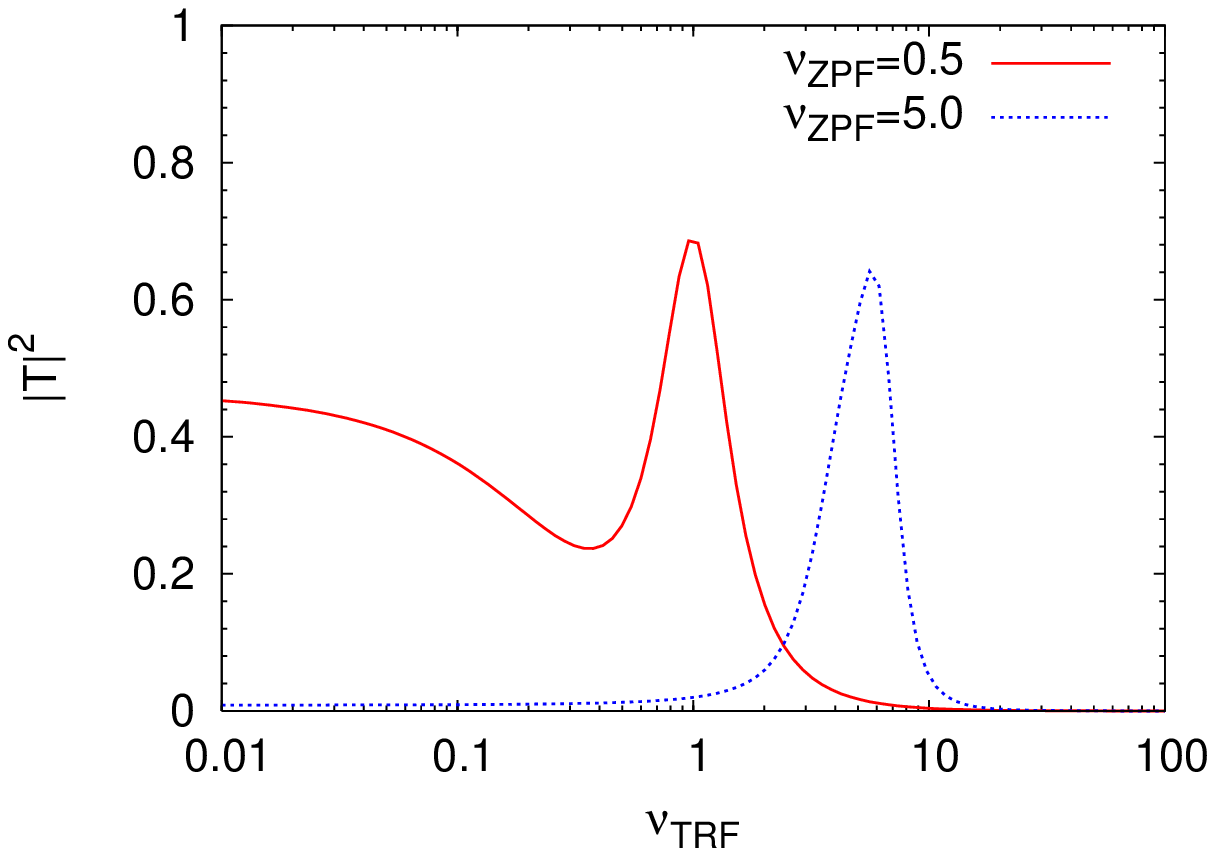} \\
\includegraphics[clip,width=0.75\hsize]{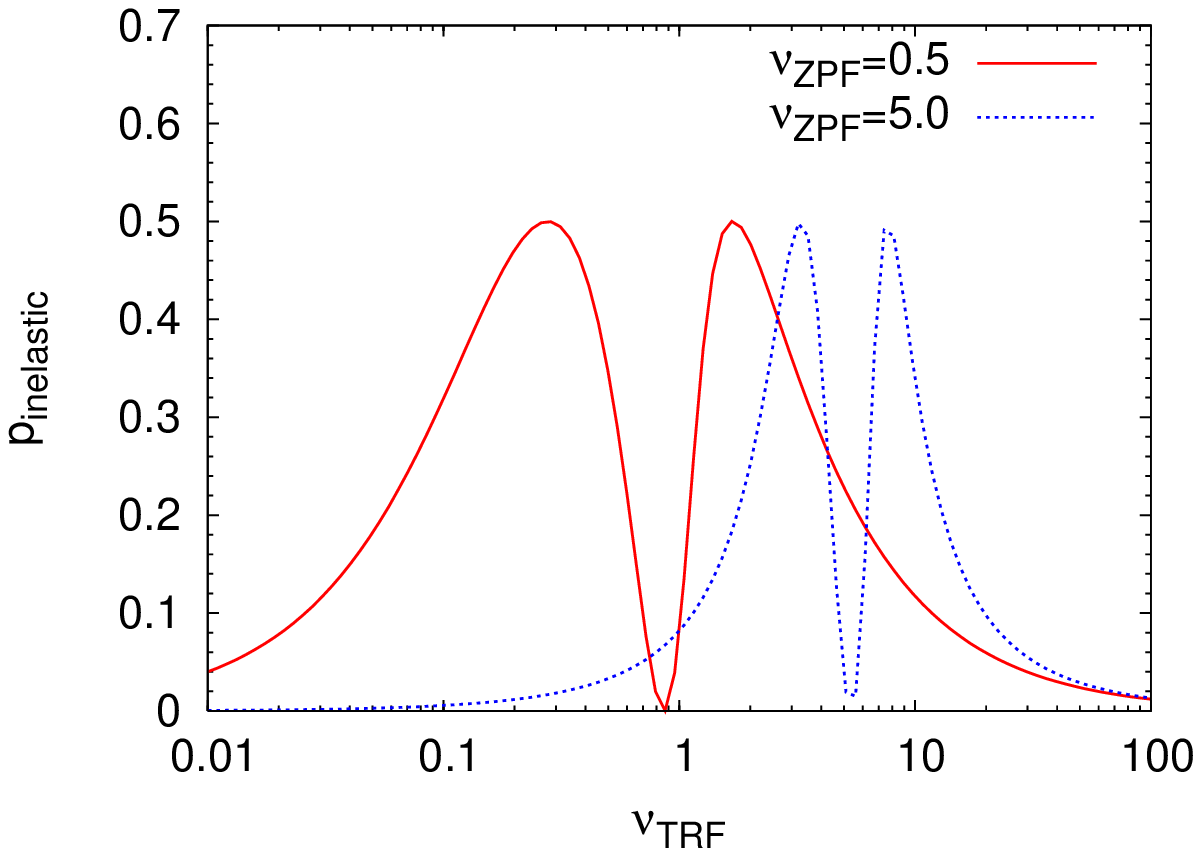}
\end{center}

{\footnotesize {\bf Fig.5:} 
The transmission $|\mathcal{T}|^2$ (upper panel) 
and the inelastic cross section $p_{\tbox{inelastic}}$ (lower panel)
versus ${\nu_{\tbox{TRF}}}$ for the simple model of Section~7 
with the same parameters as in Figs.~3-4.}
}

\clearpage


\mpg{
\begin{center}
\includegraphics[clip,width=0.9\hsize]{CeffVsNuZ0c0fin} \\
\includegraphics[clip,width=0.9\hsize]{LmbdVsNuZ0c0fin}
\end{center}

{\footnotesize {\bf Fig.6:} 
Plots of ${c_0}$ and $\lambda_0$ versus ${\nu_{\tbox{ZPF}}}$. 
For sake of comparison we also plot $c_{\tbox{eff}}$ and 
${ \lambda \equiv (\nu_{\tbox{eff}} / \nu_{\tbox{TRF}})^{1/2} }$
for two non-zero values of ${\nu_{\tbox{TRF}}}$.
Here and in the next figure we consider a bath that consists 
of $7$~TRF scatterers with ${\omega_{\alpha} \sim 0.0003}$,  
and $4$~ZPF scatterers with ${\omega_{\alpha}\sim 0.96}$. 
There is no static scatterer (${c_{\tbox{S}}=0}$).  
The kinetic energy of the incident particle is ${\epsilon_k=0.6}$.} 
}

\ \\

\mpg{
\begin{center}
\includegraphics[clip,width=0.9\hsize]{NeffVsNuTc0fin} \\
\includegraphics[clip,width=0.9\hsize]{PinVsNuTc0fin}
\end{center}

{\footnotesize {\bf Fig.7:} 
Plots of ${\nu_{\tbox{eff}}}$ and $p_{\tbox{inelastic}}$ versus ${\nu_{\tbox{TRF}}}$ 
for the same bath as in the previous figure.} 
}

\ \\

\mpg{
\begin{center}
\includegraphics[clip,width=0.9\hsize]{PinVsTeffc0fin}
\end{center}

{\footnotesize {\bf Fig.8:} 
The inelastic cross section $p_{\tbox{inelastic}}$ 
versus the scaled intensity of the thermal fluctuations $|\mathcal{T}|^4\nu_{\tbox{eff}}$,  
using the data points of Fig.~7.} 
}

\clearpage

\end{document}